\documentclass{article}

\usepackage{PRIMEarxiv}

\usepackage[utf8]{inputenc} 
\usepackage[T1]{fontenc}    
\usepackage{hyperref}       
\usepackage{url}            
\usepackage{booktabs}       
\usepackage{amsfonts}       
\usepackage{nicefrac}       
\usepackage{microtype}      
\usepackage{lipsum}
\usepackage{fancyhdr}       
\usepackage{graphicx}       
\graphicspath{{media/}}     

\usepackage{rotating}
\usepackage{array}
\newcolumntype{L}[1]{>{\raggedright\let\newline\\\arraybackslash\hspace{0pt}}m{#1}}
\newcolumntype{C}[1]{>{\centering\let\newline\\\arraybackslash\hspace{0pt}}m{#1}}
\newcolumntype{R}[1]{>{\raggedleft\let\newline\\\arraybackslash\hspace{0pt}}m{#1}}

\title{Secure human oversight of AI: \\ Threat modeling in a socio-technical context
}

\author{
  Jonas C. Ditz$^{*}$, Veronika Lazar$^{*}$, Elmar Lichtmeß, Carola Plesch, Matthias Heck \\
  Federal Office for Information Security \\
  Bonn \\
  Germany\\
  \texttt{veronika.lazar@bsi.bund.de} \\
  $^{*}$Both authors contributed equally to this research.
   \And
  Kevin Baum \\
  German Research Center for Artificial Intelligence \\
  Saarbr\"{u}cken \\
  Germany \\
  \texttt{Kevin.Baum@dfki.de} \\
   \And
  Markus Langer \\
  University of Freiburg \\
  Freiburg \\
  Germany \\
  \texttt{markus.langer@psychologie.uni-freiburg.de} \\
}

\begin{document}
\maketitle

\begin{abstract}
Human oversight of AI is promoted as a safeguard against risks such as inaccurate outputs, system malfunctions, or violations of fundamental rights, and is mandated in regulation like the European AI Act. Yet debates on human oversight have largely focused on its effectiveness, while overlooking a critical dimension: the security of human oversight. We argue that human oversight creates a new attack surface within the safety, security, and accountability architecture of AI operations. Drawing on cybersecurity perspectives, we model human oversight as an IT application for the purpose of systematic threat modeling of the human oversight process. Threat modeling allows us to identify security risks within human oversight and points towards possible mitigation strategies. Our contributions are: (1) introducing a security perspective on human oversight, (2) offering researchers and practitioners guidance on how to approach their human oversight applications from a security point of view, and (3) providing a systematic overview of attack vectors and hardening strategies to enable secure human oversight of AI.
\end{abstract}


\section{Introduction}

Human oversight of artificial intelligence (AI) is often considered a strategy to address risks associated with the use of AI in high-risk contexts \cite{green2022flaws, laux_institutionalised_2024, shneiderman_dangers_2016}. Human oversight personnel are expected to detect inaccurate outputs, outputs that may pose risks to fundamental human rights (e.g., discriminatory outputs), unsafe states of AI systems, and system malfunctions, and to intervene in order to mitigate or minimize these risks \cite{langer_effective_2024}. Human oversight is also a key concept in AI governance and ethics \cite{laux_institutionalised_2024, green2022flaws} and a central requirement for the implementation of high-risk AI systems in emerging AI regulation, such as Article 14 of the European AI Act \cite{enqvist_human_2023, laux_institutionalised_2024}. However,  relying on humans to oversee AI also faces criticism and substantial challenges in practice 
\cite{jobin2019global,green2022flaws,sterz2024quest}. The most prominent criticisms and challenges include doubts about whether humans can effectively oversee AI systems \cite{green2022flaws}, the difficulty of defining what constitutes effective human oversight \cite{sterz2024quest}, clarifying who should provide oversight and when it should occur \cite{enqvist_human_2023}, specifying which risks and unwanted system behaviors oversight personnel should detect \cite{langer_effective_2024}, and determining how to design and evaluate effective human oversight \cite{sterz2024quest}.

However, one key challenge that remains underexplored is the security risks that arise from human oversight of AI. Because human oversight will be implemented as part of the safety, security, and accountability architecture of AI systems \cite{laux_institutionalised_2024, jobin2019global}, malicious actors who successfully attack elements related to oversight can significantly undermine the safety of AI operations. In this sense, the human oversight architecture creates a new attack surface for targeting AI operations in high-risk contexts such as critical infrastructure, public administration, or medicine. A security-focused perspective on human oversight is crucial, given the central role human oversight already plays, the expectations placed on it in high-risk contexts, and its growing importance as a regulatory requirement in AI governance, such as under the EU AI Act. If we fail to assess and address the security challenges of human oversight, we risk overlooking potential attack vectors in high-risk contexts; and without awareness of these vectors, we will not be able to design effective countermeasures and hardening strategies. 
Ultimately, neglecting security aspects of human oversight can undermine its main objective---risk mitigation---or even increase risks by introducing additional attack vectors.

In this paper, we introduce a security-related perspective on human oversight of AI. 
Human oversight is best understood as a socio-technical system that can be instantiated in many different ways. For the purposes of threat modeling, we model it as an IT application and draw on established cybersecurity methods to analyze attack surfaces and vulnerabilities. This enables us to identify security threats to the key requirements of effective human oversight (i.e., epistemic access, self-control, causal power, and fitting intentions; \cite{sterz2024quest}).
Threat modeling further provides insights into applicable countermeasures and hardening strategies to mitigate the identified cybersecurity risks.

This paper makes three key contributions. First, we introduce a security-related perspective on human oversight of AI. This perspective is critically missing from  research, even though human oversight is becoming an integral component of the safety, security, and accountability architecture of AI operations, and thus an increasingly viable target for malicious actors. Second, we provide guidance on how to perform threat modeling in socio-technical contexts and identify cybersecurity threats of human oversight with possible countermeasures. Thereby, this work provides a starting point for future threat modeling of implementations of human oversight. Third, we provide a list of security vulnerabilities and hardening strategies that are applicable to human oversight applications in general and should always be considered by practitioners. We hope these contributions support AI governance and system design in ensuring not only effective but also secure human oversight of AI.

\section{Related Work}

\subsection{Effective Human Oversight of AI}
Research discusses the nature of human oversight of AI, including the \textit{what}, \textit{when}, and \textit{who} of oversight. Regarding the \textit{what}, many forms of human-AI collaboration 
involve 
human oversight. Across a wide range of human-AI interaction use cases, humans monitor AI system outputs, check their accuracy, request explanations, adapt outputs, intervene when systems malfunction, or take over manual control in the event of failures \cite{lai_towards_2023, bansal_does_2021, Yurrita_2025, Ehsan_HCXAI_2022, Liao_questioningAI_2020, Tahaei_HCRAI_2023, liao2021human, Buccinca_motivation_2024}. In a more targeted sense, and following how emerging research has defined the concept \cite{sterz2024quest}, human oversight specifically refers to humans mitigating risks in AI operations \cite{mcbride_understanding_2011, enqvist_human_2023}. More concretely, oversight personnel should monitor AI systems, detect failures and malfunctions, identify unsafe states, flag inaccurate outputs, and recognize outputs that may pose risks to fundamental human rights, such as discriminatory outputs \cite{sterz2024quest, langer_effective_2024, enqvist_human_2023}. Once such issues are detected, oversight personnel are expected to act in ways that effectively mitigate risks \cite{langer_effective_2024, sterz2024quest}. This can involve taking over manual control, overwriting inaccurate or inadequate outputs \cite{Megdi2023_workwithandworkforAI}, or stopping AI operations and triggering intervention by other human stakeholders, such as developers who can adapt the system. Importantly, the AI system under oversight may be operated by another human who is not considered oversight personnel but a user of the system. For example, a patient using an AI chatbot for mental health support \cite{Wester2024_thischatbot} would be the user, while oversight personnel monitor the patient–AI interaction to detect risky situations (e.g., if the patient discusses suicidal behavior with the chatbot).

Regarding the \textit{when}, human involvement is important at every stage of AI development: during system design, at run time, and when inspecting errors and failures in hindsight \cite{shneiderman_dangers_2016, shneiderman_human-centered_2020, Gomez-Beldarrin2025_Whydoesautomation}. In a more targeted sense, human oversight usually refers to personnel overseeing AI systems during run time \cite{sterz2024quest, laux_institutionalised_2024}. 
This ranges from dynamic, time-critical situations like autonomous vehicle oversight \cite{Megdi2023_workwithandworkforAI, Kraus_2019} to less time-pressured contexts like judicial AI \cite{langer_effective_2024, Jin2024_beyondreasonabledoubt}.

Regarding the \textit{who} of human oversight, any human with the skills necessary to effectively oversee an AI system could serve as oversight personnel \cite{enqvist_human_2023, Xie2025_exploringwhatpeople}. These individuals may be domain experts such as doctors or judges, who can evaluate the adequacy of specific AI outputs based on their expertise. Alternatively, oversight personnel may be trained to detect AI failures without being domain experts in the task performed by the AI. For example, in a medical AI setting, one oversight person might monitor the system’s technical functions on an aggregate level, while medical professionals evaluate individual outputs requiring domain knowledge \cite{Xie2025_exploringwhatpeople, langer_effective_2024}. Human oversight can also be performed by a single individual or by a team collaboratively overseeing AI systems.
Research has also examined the requirements for effective human oversight. Sterz et al. \cite{sterz2024quest} identify four key aspects: epistemic access, causal power, self-control, and fitting intentions. 

Epistemic access means that oversight personnel possess sufficient knowledge about the AI system and understand the consequences of their interventions. Without it, they may fail to detect system failures or inaccurate outputs. Causal power refers to the ability to intervene in ways that influence AI processes and outputs. Without it, oversight personnel may be unable to overwrite or correct system outputs. Self-control requires oversight personnel to maintain agency over their own actions. For instance, fatigue or intoxication may impair their capacity to fulfill oversight duties. Likewise, being threatened by another person can also undermine self-control, as the actions performed by the oversight personnel are then steered by another agent. Fitting intentions means that personnel are motivated to carry out their responsibilities with the right intentions. In an extreme case, a malicious actor might be hired as oversight personnel and deliberately undermine AI safety. In a less extreme case, poor working conditions may erode motivation and attention, weakening oversight effectiveness \cite{sterz2024quest}.

Finally, regarding how to design effective human oversight, insights from the broader field of human-AI collaboration provide valuable guidance (e.g., \cite{lai_towards_2023, amershi_guidelines_2019, mcbride_understanding_2011}). Research emphasizes the sociotechnical nature of oversight and highlights the need to design technical, human, and contextual factors that support effective oversight \cite{langer_effective_2024, laux_institutionalised_2024, sterz2024quest}. Technical factors include system interface design, explainability approaches, and interaction modes between humans and AI systems \cite{bansal_does_2021, langer_what_2021, miller_explanation_2019, bucinca_trust_2021}. Human factors include the skills oversight personnel need and the training they receive \cite{Xie2025_exploringwhatpeople, mcbride_understanding_2011, langer_effective_2024}. Contextual factors include the design of the working environment, time pressure, and the clarity of roles and responsibilities within oversight teams \cite{langer_effective_2024, Yun2025_generativeai, Parker_2020}.

\subsection{Attack Vectors and Hardening Strategies in Cybersecurity}
With the implementation of AI systems across diverse context, ensuring the security of deployment has become increasingly important. Enabling systematic threat modeling for AI-based applications requires an in-depth understanding of specific attack vectors and the attack surface within an application domain \cite{theisen2018attack}. Traditionally, an attack vector is defined as a method or pathway that a malicious actor uses to gain access to a network or system. In this paper, we extend the definition to also include methods or pathways that undermine the effectiveness of human oversight. We use the term attack surface to describe the total set of points where a malicious actor could attempt to enter, exploit, or otherwise attack digital and AI infrastructures \cite{theisen2018attack}. While there is research that systematically analyses AI-based systems from a cybersecurity point of view \cite{gao2021autonomous,qammar2022federated}, such systematic investigations are still missing for human oversight.

Related work can be found in the field of AI cybersecurity, which examines attack vectors that explicitly target AI systems. For instance, poisoning attacks manipulate training data to introduce backdoors into AI models that can later be exploited \cite{shafahi2018poison,gu2019badnets,jagielski2021subpopulation}. Adversarial attacks embed imperceptible patterns into inputs to change model behavior, often causing misclassifications or other problematic outcomes \cite{huang2017adversarial,zugner2018adversarial,akhtar2018threat}. Another line of research targets explainable AI (XAI) methods designed to enhance understanding of AI decisions \cite{langer_what_2021}. Such attacks manipulate explanations to obscure the real reasons behind a model’s decision \cite{slack2020fooling,aivodji2019fairwashing,slack2021counterfactual}. However, AI system are still IT systems at the core. This means that traditional attack vectors, like distributed denial-of-services (DDOS) \cite{lau2000distributed,paxson2001analysis}, man-in-the-middle (MITM) \cite{conti2016survey}, or social engineering attacks \cite{alizadeh2023catch,distler2023influence}, have to be considered for AI systems as well. Furthermore, human-centered attack vectors that are usually underexplored in cybersecurity, e.g., coercion, bribery, and insider threats like disgruntled employees \cite{holovkin2021corruption,weber2017coercion,maasberg2020dark}, are of special importance for human oversight. 

In response to the many ways digital systems can be attacked, cybersecurity research has developed hardening strategies as methods designed to secure AI systems against attacks \cite{ross2019developing,reda2023taxonomy,banik2023automated}. One line of work establishes methods for securing data traffic in computer networks. Network management systems can detect malicious traffic \cite{stallings1995network,iitumba2023reconsidering}, while encryption protects legitimate communications \cite{menezes2018handbook,al2024impact}. To monitor organizational IT systems, intrusion detection systems have been introduced and are continuously improved \cite{khraisat2019survey,kheddar2025transformers}. Another widely used practice is red teaming, where every system component is rigorously tested by simulating attacks; research continues to optimize red teaming approaches \cite{oakley2019professional,zhang2024human}. Beyond technical strategies, organizational measures also play a key role. Transparency of software components is crucial for analyzing systems and identifying potentially harmful functionalities \cite{hughes2023software,hall2025pitfalls}. Other well-established approaches include governance principles and audit frameworks that help secure deployed systems \cite{sabillon2017comprehensive,falco2021governing}. Finally, like attack vectors, hardening strategies must also address the human factor. Research has explored how to improve system security by training users and raising awareness of risks \cite{glas2023train,oh2025understanding}. 

In this paper, we extend the literature by contextualizing attack vectors and hardening strategies specifically for securing human oversight of AI and show how to use this knowledge for systematic threat modeling of human oversight.

\section{Threat modeling of human oversight}\label{sec:threat_modeling}

Human oversight can serve as part of the safety, security, and accountability architecture of AI systems, as oversight personnel are expected to detect and mitigate risks. In addition, oversight personnel may be required to actively confirm that an AI system is in a safe state. Human oversight may become even more important in AI operations if it evolves into a requirement of AI governance and regulation in high-risk contexts \cite{green2022flaws, enqvist_human_2023, laux_institutionalised_2024, sterz2024quest}.

With a more wide-spread implementation, human oversight will also become an increasingly viable target for malicious actors seeking to corrupt or disrupt AI-based operations. Such actors could be cybercriminals targeting high-risk AI systems for ransom or foreign governments attacking a country’s critical infrastructure in hybrid or open conflict. Furthermore, sufficiently powerful AI agents that are successfully attacked could themselves become malicious actors by attempting to evade oversight \cite{amodei2016concrete,meinke2024frontier}. Considering these threats, designing human oversight requires not only ensuring effective oversight \cite{sterz2024quest} but also ensuring secure human oversight.

In cybersecurity, the standard approach to identify, quantify and address potential flaws and vulnerabilities within an IT system is to perform threat modeling. Organizations and IT-professionals can use threat modeling to examine the IT system at hand from the viewpoint of malicious actors using a structured process. 
Threat modeling consists of four steps.

First, the scope of the tested application has to be determined, usually by creating a data flow diagram (DFD).
Furthermore, the first step identifies entry points into the application, exit points of the application, privilege or trust boundaries, and assets that need to be protected. The second step involves determining threats to the tested application. In this work, the STRIDE \cite{kohnfelder1999threats} model is used for determining threats. STRIDE---an acronym for \textit{Spoofing}, \textit{Tampering}, \textit{Repudiation}, \textit{Information disclosure}, \textit{Denial of service}, \textit{Elevation of privilege}---is a well-established threat model. During the third step of threat modeling, countermeasures and potential hardening strategies are determined for identified threats. The fourth step consists of assessing the results of the previous steps and ensuring that no important points have been missed. For this manuscript, step four will be implicitly addressed in the discussion.

This section details the decisions we took in performing threat modeling for an IT-application representation of human oversight, and how we incorporated its socio-technical context into the analysis.
We provide a guideline for researchers and practitioners who want to perform threat modeling for their real-world implementations. We provide a table that summarizes the main findings of this paper in Appendix \ref{app:table}. Furthermore, Appendix \ref{app:example} provides a running example that maps our findings onto a concrete use case.

\subsection{Step 1 - The scope of human oversight}

\begin{figure}
    \centering
    \includegraphics[width=0.8\linewidth]{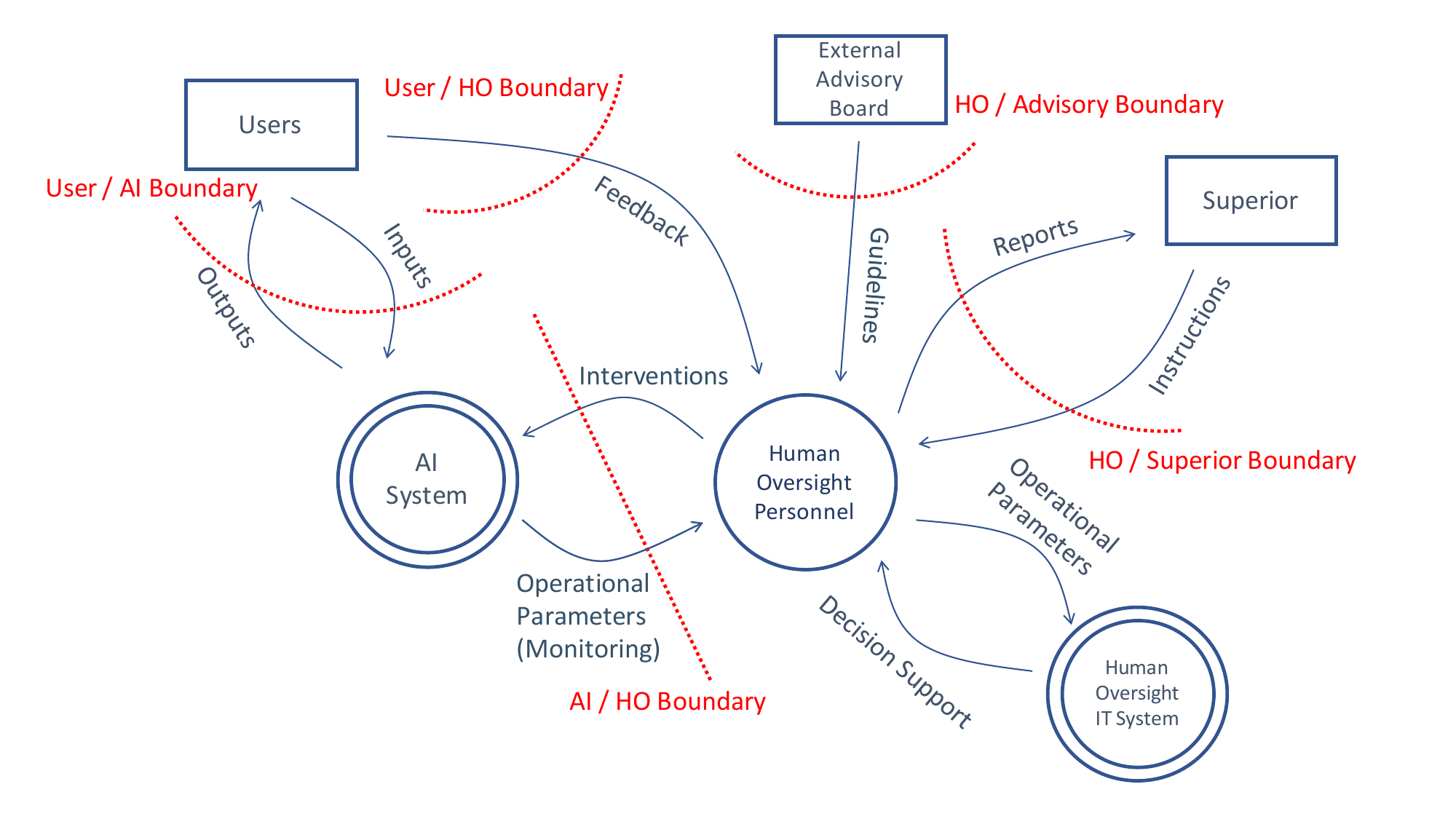}
    \caption{
    Data flow diagram (DFD) of the abstracted human oversight application used for threat modeling.
    Arrows indicate the directed paths of data through the application. Each arrow is labeled with the type of data that is transported. Circles indicate processes, two concentric circles indicate multiple subprocesses, and rectangles indicate external entities. Privilege boundaries are drawn with red dashed lines. HO = Human Oversight.}
    \label{fig:dfd}
\end{figure}

As mentioned above, the first step of any threat modeling is to collect information about the subject of the threat modeling, and thereby determine the scope of the work. A data flow diagram (DFD) shows different paths through an application and indicates the locations of privilege boundaries (also called trust boundaries). These boundaries indicate that movement along the path leads to a change of privileges, i.e., users and processes beyond the boundary have more rights to perform actions within the system like write data or run software. We extend the meaning of privilege boundaries to also include the right for intervention, meaning that entities with higher privilege can intervene on entities with lower privilege. The shape of nodes in a DFD are associated with specific meanings. A rectangle indicates an external entity, like a user that accesses the application. A circle indicates a single process. Two concentric circles indicate multiple subprocesses. Data flows are indicated by arrows, and privilege boundaries are indicated by dashed lines. Traditionally, DFDs can indicate data storage nodes with two parallel lines but we did not need this node type for the human oversight DFD.

Furthermore, DFDs are hierarchical in nature. This means that several different levels of DFDs can be created for an application. Each DFD focuses on a different part of the application. This approach has the advantage that every isolated DFD remains relatively simple. We use this property to compensate for the fact that we are not looking at a concrete implementation but a more abstract concept. Designing a high-level DFD allows us to omit implementation details via the two types of process nodes. Real-world implementations of human oversight can than fill in the details of the processes with lower-level DFDs.

\subsubsection{The human oversight DFD}
Figure \ref{fig:dfd} presents our DFD of the human oversight application. We identified three external entities. The \textit{users} node includes all entities that interact with the AI system, either humans or other IT systems. The \textit{superior} node indicates structures within the organization that have greater authority than the human oversight system. That could be for example the management of the organization. The \textit{external advisory board} node indicates an external entity that defines appropriate actions and goals of human oversight, similar to an ethical advisory boards. Next, there are two multiple processes nodes in the DFD. One node represents the AI system, the other node represents the human oversight IT system. The human oversight IT system consists of all software that the human oversight personnel need to perform their duties. By choosing multiple processes nodes for the AI system and the human oversight IT system, we can create a top-level DFD that is valid for numerous implementations of human oversight. In case of threat modeling a concrete human oversight application in the future, these multiple processes nodes simply have to be specified with additional lower-level DFDs. Finally, the last identified node represents the human oversight personnel. We model this component of the human oversight application as a process to indicate the active role of humans in human oversight. Furthermore, the crucial role of human oversight personnel to the application is shown by the high connectivity of this node. 

We identified several data flows within our theoretical human oversight application. Users send inputs to the AI system, which responds by sending outputs back to the users. Inputs can be any data that is used to interact with the AI system. Similarly, outputs can be arbitrary data that represents the output of the AI system (e.g., decisions or generated data points). There is a data flow between users and HO personnel indicating the possibility of users to directly send feedback and reports to the HO personnel. Such user reports are an important oversight signal in many applications. Another data flow connects the AI system with the human oversight personnel. While the AI system send operational parameters to the human oversight personnel (this is often called monitoring in the HO literature), the oversight personnel performs interventions on the AI system \cite{langer_effective_2024}. These can include commands to shut down, corrections of model parameters, or other interventions. The next data flow is located between human oversight personnel and human oversight IT system. While the personnel sends the operational parameter from the AI system to the human oversight IT system, the IT system uses these parameters to provide decision support for the oversight task. These will be transferred back to the human oversight personnel. The operational parameters could be directly send from the AI system to the human oversight IT system. However, This would not create a new privilege boundary crossing and, thus, can be viewed as similar to our DFD. We decided on the current DFD design to emphasis the importance of the human oversight personnel. There is a data flow between the human oversight personnel and the superior node. Since the superior node represents an entity of higher authority, the data flow labeling is chosen to represent this hierarchy. The human oversight personnel sends reports to the entity with higher authority while instructions (e.g., work instructions) are flowing the other way. Finally, data in form of guidelines flow from the external advisory board to the human oversight personnel.

Five privilege boundaries can be identified in the DFD. First, there is a privilege boundary between users and the AI system. While users should have minimal rights that are necessary to use the application, the AI system needs enough rights to fulfill its task. Additionally, there is a privilege boundary between users and HO personnel with a similar reasoning. The next privilege boundary lies between the AI system and the human oversight personnel. Being able to intervene on the AI system requires the HO system to have higher privileges that the AI system. The third privilege boundary lies between the human oversight personnel and the superior node. Since the superior node represents management of an organization, this node has the higher privilege context. The fourth privilege boundary lies between the human oversight personnel and the external advisory board. An external advisory board has no privileges within the system. That means we have five different privilege levels: advisory privilege, user privilege, AI privilege, HO privilege, and superior privilege.

\subsubsection{Entry points}

\begin{table}[]
    \centering
    \caption{A list of all identified entry points into the human oversight application. Each entry point has a unique identifier, a name, and a description. Furthermore, the column \textit{privilege levels} indicate which privilege level are required at each entry point.}
    \small
    \begin{tabular}{l|l|L{8cm}|l}
        \toprule
        \textbf{ID} & \textbf{Entry point} & \textbf{Description} & \textbf{Privilege levels}  \\
        \midrule
        Entry.1 & User login & Login procedure for users to interact with the AI system. This could be a login page reachable from the internet using a browser. Other possibilities are APIs, specialized user applications, or hardware components. & user, AI, HO, superior \\
        \midrule
        Entry.2 & HO personnel login & Login procedure for human oversight personnel to get access to the human oversight system. Possible options for this entry point are browser-based login pages, specialized applications, or hardware components. & HO, superior \\
        \midrule
        Entry.3 & Access of superiors & Capabilities of entities with higher authority (e.g., management of organizations) to gain access to the human oversight system. This can include organizational structures and intervention capabilities. & superior \\
        \bottomrule
    \end{tabular}
    \label{tab:entry_points}
\end{table}

The next part of determining the scope of human oversight is to identify entry points into the application. The identification of entry points plays a crucial role in understanding possible paths that are open to a malicious actor to attack an application. Usually, this means to determine technical entry points like HTTP ports, login pages or functions, or search entry pages. Since we are not looking at a concrete implementation of human oversight, our analysis focuses on potential points of interest that need to be closely examined for real-world applications of human oversight. For every entry point, we also specify which privilege levels are required to use the entry point. Table \ref{tab:entry_points} shows a summary of the entry points. The three identified entry points are the component used by users to access and use the AI system, the component used by human oversight personnel to access the human oversight application, and the component that allows entities with higher authority to access the human oversight application.

\subsubsection{Exit points}

\begin{table}[]
    \centering
    \caption{A list of identified exit points of the human oversight application. Each exit point has a unique identifier, a name, and a description. The last column indicates the corresponding entry point to each exit point, if such an entry point exists.}
    \small
    \begin{tabular}{l|l|L{8cm}|l}
        \toprule
        \textbf{ID} & \textbf{Exit point} & \textbf{Description} & \textbf{Corr. entry point} \\
        \midrule
        Exit.1 & AI Output & Communication channel that is used by the AI system to deliver outputs (e.g., predictions, decision, or generated data points) to users. & Entry.1 \\
        \midrule
        Exit.2 & HO authentication & Response functionality by the human oversight application to deliver the results of the human oversight personnel's login procedure. & Entry.2 \\
        \midrule
        Exit.3 & HO system report & Communication channel between human oversight system and entities with higher authorities. This channel is used to deliver activity report and other requested information to superiors like the management of an organization. & Entry.3 \\
        \bottomrule
    \end{tabular}
    \label{tab:exit_points}
\end{table}

Some attacks require an exit point to achieve their goal, i.e., a gateway that can be used to ex-filtrate data from the application. Two more commonly known attacks worth mentioning here are cross-site-scripting vulnerabilities and information disclosure vulnerabilities. In general, exit points that are not sufficiently secured can be exploited by attackers to collect sensitive information. From a technical point of view, exit points often have a corresponding entry point. All exit points are summarized in Table \ref{tab:exit_points}. We identified three exit points for the human oversight application: the AI output provided to users, the component allowing human oversight personnel to access the human oversight application, and the component used to provide work reports to entities with higher authority.

\subsubsection{Assets}

\begin{table}[]
    \centering
    \caption{List of identified physical assets of the human oversight application. Each asset has a unique identifier, a name, and a description.}
    \small
    \begin{tabular}{l|l|L{9cm}}
        \toprule
        \textbf{ID} & \textbf{Name} & \textbf{Description} \\
        \midrule
        PA.1 & HO personnel credentials & The keys that human oversight personnel uses to access the human oversight system. \\
        \midrule
        PA.2 & User data & Depending on the concrete application, the human oversight system might store sensitive information about users, if this information is needed to determine whether an intervention has to be performed. \\
        \bottomrule 
    \end{tabular}
    \label{tab:physical-assets}
\end{table}

To complete the first step of threat modeling human oversight, we identify assets that need to be protected. Assets can be both physical and abstract. Physical assets are for example sensitive information like medical data of patients. Abstract assets can be understood as important properties and states, like the reputation of an organization. The lists of physical and abstract assets presented in this work are not exhaustive. Real-world applications of human oversight are likely to have additional assets. 

\begin{table}[]
    \centering
    \caption{List of identified abstract assets of the human oversight application. Each asset has a unique identifier, a name, and a description.}
    \small
    \begin{tabular}{l|l|L{10cm}}
        \toprule
        \textbf{ID} & \textbf{Name} & \textbf{Description} \\
        \midrule
        AA.1 & Confidentiality & Ensuring that sensitive information about the human oversight application, like interventions or decision protocols, cannot be accessed by unauthorized individuals. \\
        \midrule
        AA.2 & Integrity & Ensuring the correctness of data and functionality of the human oversight application. \\
        \midrule
        AA.3 & Availability & Ensuring that the human oversight application is functional at all times. \\
        \midrule
        AA.4 & Epistemic access & Ensuring that human oversight personnel can sufficiently understand how the AI system works. \\
        \midrule
        AA.5 & Causal power & Ensuring that the human oversight system can sufficiently intervene into the AI system. \\
        \midrule
        AA.6 & Self-control & Ensuring that the human oversight personnel has sufficient control over their own actions. \\
        \midrule
        AA.7 & Fitting intentions & Ensuring that the internal goals of human oversight personnel is aligned with the goals of the human oversight application. \\
        \bottomrule
    \end{tabular}
    \label{tab:abstract-assets}
\end{table}

We identify two physical assets (Table \ref{tab:physical-assets}). The first are the credentials of human oversight personnel that are used to log into the human oversight system. These would allow an attacker to gain access with high privileges. Another physical assets are sensitive user information. Depending on the AI system, additional information about users might be necessary to ensure that the human oversight system has sufficient knowledge about a situation to decide whether or not to intervene. Envision an AI system that acts as a medical decision support system. Judging the validity of decisions made by such a system requires detailed knowledge of patient-specific information. In general, such sensitive information has to be protected against unauthorized access.

We identified seven abstract assets (Table \ref{tab:abstract-assets}). The first three assets that need to be protected are defined by the basic focus of cybersecurity, the CIA triad. This concept describes three main aspects of IT systems that have to be protected. The first aspect is the confidentiality of information within the system (denoted by the letter 'C'). Confidentiality is closely connected with the physical assets identified above but also covers appropriate management of read access within the application. The second aspect is the integrity of data and functionality of the IT system (denoted by the letter 'I'). For human oversight this can include for example ensuring correctness of AI operational parameters or correct functioning of software components. The last aspect covered by the CIA triad is availability (denoted by the letter 'A'). For example, it needs to be ensure that the human oversight personnel can access the AI system or the human oversight IT system. Furthermore, the obligations associated with human oversight can only be fulfilled, if such oversight can be considered effective. Therefore, the four requirements of effective oversight \cite{sterz2024quest} have to be added to the list of abstract assets. The first asset is epistemic access, which means that human oversight personnel has to be able to understand the functioning of the AI system. The next asset is causal power, which means that the human oversight system has to be able to perform interventions on the AI system. The third requirement is self-control, which means that the human oversight personnel of our application can make decisions without malicious outside influence. Finally, the fourth requirement is fitting intentions. Mapping this requirement onto our human oversight application means that the goals of the human oversight personnel have to be aligned with the overall goals of human oversight.

Based on the result of the first step of threat modeling, we now continue to identify security threats that need to be considered when implementing a human oversight application.

\subsection{Step 2 - Security threats}\label{sec:stride}
We use the established threat model STRIDE (Spoofing, Tampering, Repudiation, Information disclosure, Denial of services, Elevation of privilege) to identify security threats to the human oversight application. STRIDE allows us to examine the application from an attacker's perspective and to identify vulnerabilities with the attacker's goals in mind.

\subsubsection{Spoofing}
The term spoofing summarizes threat actions that are aimed at accessing and using credentials without permission. The goal is to interact with a system while assuming a false identity. This can allow attackers to gain unauthorized access to sensitive data or functionalities.

The obvious target for spoofing attacks in our human oversight application is the human oversight personnel and, thereby, using entry point \textit{HO personnel login} (Entry.2) for exploitation. Human oversight personnel has to have far reaching privileges within the human oversight application, thus, gaining access through the HO personnel login can be a high priority target for attackers. The first set of tools readily available to attackers to pursue this target are the different methods from social engineering. This include water holing, baiting, phishing, vishing, smishing, and others. Any of the mentioned approaches can be used to steal credentials and use them for unauthorized access to the human oversight application with considerable privileges. If there is insufficiently secured remote access to the human oversight application, attackers can perform a man-in-the-middle (MITM) attack to get transferred information, which in this case are credentials. The term MITM describes attacks that rely on the attacker's ability to secretly intercept, relay, and possibly alter communication between two parties. The abstract assets that are vulnerable to social engineering attacks are \textit{confidentiality} (AA.1) and \textit{self-control} (AA.6). The vulnerability of confidentiality is self-evident. Since the credentials themselves are considered confidential information, attacker access to credentials is a breach of confidentiality. The vulnerability of self-control becomes clear when looking at the goal of spoofing attacks: accessing an application while assuming a false identity. Attackers that use stolen credentials and are able to use these credentials to log into the human oversight application can act as human oversight personnel. For the system, all malicious actions taken will look like actions by human oversight personnel. Therefore, the human oversight personnel is no longer in control of their own actions. The physical asset \textit{HO personnel credentials} (PA.1) is a target of these types of attacks.

Another possible direction of spoofing attacks has emerged with the rise of more and more powerful agentic AI. This direction comes from the AI agents themselves. Research has suggested that current state-of-the-art large language models (LLMs) have the ability to do scheming \cite{meinke2024frontier}. This term summarizes different misaligned behavioral patterns shown by LLMs that are meant to pursue objectives that differ from expectations of human users. With insufficiently secured model context protocol (MCP) servers, AI agents can try and use available tools to access credentials of human oversight personnel. Using these credentials, the AI agent can act within the human oversight system acting as the human oversight personnel. In this scenario, human oversight would be completely bypassed by the AI agent. Obviously, one abstract asset that is attacked in this scenario is \textit{integrity} (AA.2). An agentic AI bypassing human oversight renders the functionality of the human oversight application useless. Furthermore, \textit{self-control} (AA.6) is also attacked since a successful attack results in another entity being in control of the human oversight personnel's actions. One might argue that also the assets \textit{confidentiality} (AA.1) and \textit{HO personnel credentials} (PA.1) are attacked. However, we would argue that misaligned behavior of AI agents do not expose confident data to the outside of the application. Therefore, these assets are not attacked.

\subsubsection{Tampering}
The term tampering summarizes attacks that are aimed at maliciously modifying a system. This can include changing persistent data, altering the system's behavior, or messing with communication in transit.

One target for tampering attacks is the communication from users to the AI system. Malicious actors can use (il)legitimately obtained user credentials to send (indirect) prompt injections to the AI system. Additionally, MITM attacks can be used to modify legitimate traffic coming from users with (indirect) prompt injections. Such attacks can cause AI systems to produce unexpected behavioral patterns. First asset vulnerable to this attack is \textit{integrity} (AA.2) since (indirect) prompt injection cause unexpected behavior of AI systems and, thus, attack the functional integrity of the system. The second vulnerable asset is \textit{epistemic access} (AA.4). Since tampering attacks alter the behavior of AI systems and can go undetected, the knowledge of the HO system on how the AI system works gets eroded. If the AI system has the rights to write data, the physical asset \textit{user data} (PA.2) can also be vulnerable to these attacks. Malicious actors can try to exploit these rights and trigger the AI system to change stored user data.

Another possible tampering attack is targeting the supply chain to poison training data or directly modify model parameters. Poisoning attacks can embed back doors into AI models, which can be exploited once the model is deployed. Again, backdooring AI models and tampering with parameters erode the HO system's fundamental understanding of how the AI system functions. Thus, \textit{epistemic access} (AA.4) is a target of these attacks. Furthermore, \textit{integrity} (AA.2) is also a target, since the attacks aim to compromise the functional integrity of the AI system. 

\subsubsection{Repudiation}
The term repudiation summarizes attacks that are aimed at performing prohibited operations within an IT system. These can exploit lacking capabilities to track operations within a system. 

The first attack vector that can be exploited for repudiation are adversarial attacks and (indirect) prompt injections. Using the communication channel between AI system and users, malicious actors can try to circumvent guardrails embedded into AI systems to trigger forbidden operations. In the worst case, sensitive data can be extracted exploiting the lethal trifecta \cite{willison2025lethal} together with exit point \textit{AI output} (Exit.1). If the AI system has access to log files, malicious actors can try to hide these forbidden operations. Therefore, (indirect) prompt injections can be used to attack both physical assets, \textit{HO personnel credentials} (PA.1) and \textit{User data} (PA.2). Furthermore, the abstract assets \textit{confidentiality} (AA.1), \textit{integrity} (AA.2), and \textit{epistemic access} (AA.4) are vulnerable to these attacks.

LLM's ability to perform scheming can also open the human oversight application to repudiation attacks. Scheming can convert an AI system into an insider threat by triggering forbidden operations due to misaligned behavior. Again, these attacks can be obscured, if the AI system has the ability to alter log files. \textit{Epistemic access} (AA.4) and \textit{integrity}(AA.2) are vulnerable to this form of attack, since scheming prevents human oversight personnel to fully understand the functioning of the AI system and the functionality of the AI system is compromised. Depending on how extensive the misaligned AI system can operate within the application, the asset \textit{causal power} can also be vulnerable. An AI system with sufficient rights can potentially limit the intervention capabilities of the human oversight personnel.

Another set of attack vectors for repudiation attacks are directly targeted at the human oversight personnel. Bribery and coercion can be used by malicious actors to force human oversight personnel to perform prohibited operations. Again, being able to alter the log files allows human oversight personnel to hide these forbidden operations. The abstract assets \textit{confidentiality} (AA.1) and \textit{integrity} (AA.2) are vulnerable to these types of attacks, since both bribery and coercion can be used to extract sensitive information from the system and maliciously alter persistent data and functionalities. Furthermore, coercion attacks \textit{self-control} (AA.6) due to the outside force that determines the actions for the human oversight personnel. Bribery attacks \textit{fitting intentions} (AA.7), since the original goal of the human oversight personnel gets overridden with monetary or otherwise self-serving interests.

\subsubsection{Information disclosure}
The term information disclosure summarizes attacks that aim at reading files without permission or reading data in transit. Since all attacks that fall into this category are explicitly targeting the disclosure of sensitive information, the asset that is mainly targeted by these attacks is \textit{confidentiality} (AA.1). Furthermore, the two identified physical assets \textit{HO personnel credentials} (PA.1) and \textit{user data} (PA.2) are sensitive information and, thus, can also be targeted by information disclosure attacks.

The two data flows that will be the main target of information disclosure attacks are between AI and human oversight personnel, and human oversight personnel and IT system. Both data flows contain valuable information about the inner workings of the human oversight applications and, hence, might be interesting for attackers. The data flow between users and AI system could also be attacked. However, this data flow contains mainly information about the inner workings of the AI system. This data flow plays only a secondary role for the threat modeling of human oversight. A standard attack vector for data in transit are MITM attacks.

\subsubsection{Denial of services}\label{sec:denial}
The term denial of services covers attacks attempting to deny access to valid users, such as by making a web server temporarily unavailable or unusable. One of the most well-known attack patterns that fall into this category is flooding a server with more requests than can be reasonably processed. This leads to the server becoming unresponsive. These attack patterns are also called (distributed) denial of services ((D)DOS) attacks. In general, threat actions that fall under the term denial of services are targeting the abstract asset \textit{availability} (AA.3).

The first obvious target for denial of services attacks is the human oversight IT system. If the network that runs the IT system is accessible from the internet, malicious actors can deploy (D)DOS attacks to flood the network with requests. Another possibility is the exploitation of software vulnerabilities to cause a shutdown of the human oversight IT system. Since the human oversight IT system provides help to human oversight personnel to understand the AI system's behavior and decision support, a denial of service attack on the IT system additionally targets the abstract asset \textit{epistemic access} (AA.4). 

In a distributed setting, where the HO system is not co-located with the AI system, malicious actors can specifically target the network of the human oversight personnel. This would prevent human oversight personnel from intervening into the AI system and, thus, additionally targets the abstract asset \textit{causal power} (AA.5).

\subsubsection{Elevation of privilege}
The term elevation of privilege summarizes attacks that aim at escalating the own privileges within a system. An elevation of privileges attack uses vulnerabilities and design/configuration flaws after an initial network breach to get access to data and functionalities that are normally inaccessible for the initial privilege group. Due to our extended definition of privilege, an elevation of privilege attack can also lead to intervention rights that normally would not be accessible. 

An attack vector that can be exploited by malicious actors to attack the AI system within the human oversight application is jailbreaking. Here, different techniques like poisoning attacks and adversarial attack can be used to circumvent guardrails that prevent AI systems from misaligned behavior. Together with the exploitation of software and design flaws this can lead to the AI system having similar privileges than the human oversight components. In this case, interventions issued by the human oversight personnel could be overwritten by the AI system. Therefore, these attacks target the abstract asset \textit{causal power} (AA.5)

\subsection{Step 3 - Countermeasures and hardening strategies}

The STRIDE analysis revealed several vulnerabilities within the human oversight application. The next step in our threat modeling highlights a few countermeasure and mitigation strategies that should be adhered to when implementing human oversight. The list is not exhaustive since our theoretical threat modeling cannot determine all threats that an actual implementation might face. For example, there might be mitigation strategies that depend on specific hard- and software components installed into the human oversight application. Therefore, the provided list can be understood as a starting point and minimal set.

\subsubsection{Intrusion Detection System}
Several threat actions identified in section \ref{sec:stride} require an initial breach of the application environment. Therefore, the application's environment, including all computer and network components, should be closely monitored using intrusion detection systems (IDS). These systems typically combine multiple components and methods to perform this monitoring. The core technical components of an IDS include network or host sensors (depending on the IDS type), a database, a management module, and an analysis module. Host and network sensors monitor the system and collect the necessary information needed for the detection of attacks. The database needs to ensure that the data collected by sensors is stored with a sufficient degree of detail and for a sufficient amount of time. The management module provides all functionalities needed for configuring and calibrating the IDS. The analysis module provides all functionalities necessary for analyzing stored data and performing incident detection. A well-designed and maintained IDS can significantly enhance the cybersecurity of IT systems \cite{bace2001intrusion,khraisat2019survey}.

\subsubsection{Encryption}
Another mitigation strategy that protects sensitive information is encryption. Non-encrypted data traffic enables malicious actors to carry out for example Man-in-the-Middle (MitM) attacks with relative ease, often without requiring technical expertise or expensive software and hardware. This makes an application vulnerable to spoofing, tampering, and information disclosure threats. To ensure adequate security, modern cryptographic methods with strong security guarantees must be employed to encrypt data traffic \cite{kopal2014cryptool,bsi2025rypto}. Furthermore, encrypted data storage should be considered throughout the human oversight application.

\subsubsection{Network Management}
As indicated in section \ref{sec:denial}, denial of service threats overwhelm a network by generating excessive traffic. Defending against these forms of attacks requires a multi-step approach \cite{osanaiye2016distributed,bawany2017ddos}. The first step is to deploy monitoring systems that analyze network traffic and trigger alerts if a potential (D)DoS attack is detected. The second step involves methods to distinguish between benevolent traffic (typically generated by humans) and malicious traffic (often produced by bots). Finally, filtering techniques can mitigate (D)DoS attacks by restricting or limiting what users -- humans or bots -- can do within the network.

\subsubsection{Transparency}
Enhancing transparency can improve cybersecurity by facilitating the detection of attacks. While transparency itself is not typically considered a cybersecurity measure, it provides access to key security-relevant information, such as implemented components, training data, system functionality, development processes, and dependencies. When combined with other cybersecurity tools, such as vulnerability management, the information made available through transparency increases the likelihood of detecting attacks. Additionally, transparency supports auditing and compliance attestations by making security-relevant information easily accessible and enables model tracking. Transparency can also help to secure the supply chain of AI, for example with an SBOM for AI \cite{g72025sbomai}. This could help as a countermeasure against AI specific attacks, like poisoning or adversarial attacks.

\subsubsection{Training of Human Oversight Personnel}
Training aims to strengthen the resilience of human oversight personnel against attacks \cite{Schaab_2017}. Training initiatives can help personnel to recognize, resist, and properly respond to manipulation attempts. Training may include social engineering awareness, teaching personnel to identify common manipulation tactics such as phishing or impersonation of authority figures. Practical exercises, such as simulated phishing campaigns or role-play scenarios, reinforce recognition and proper reporting behaviors.

It can also address resilience against coercion, preparing personnel to respond to threats or undue pressure. This includes protocols for escalating coercion attempts, access to secure whistleblowing channels, and preparation to maintain effective human oversight under stress. Anti-bribery training can teach recognizing undue influence. Clear organizational policies defining conflicts of interest and mandatory disclosure mechanisms further reduce the risk of misaligned intentions. Training of human oversight personal counters social engineering, coercion, and bribery.

\subsubsection{Red Teaming}
The term red team refers to an independent group that investigates an organization’s IT systems to identify vulnerabilities and security flaws \cite{mansfield2018best}. The red team simulates the behavior of malicious actors, employing all methods that real attackers might use. Its goal is to uncover as many exploitable security issues within the tested system as possible. Continuous red teaming can help to stay on top of emerging security flaws after the initial threat modeling of an application. Software systems are dynamic even in deployment and, thus, require constant monitoring.

\section{Discussion}

Research on human oversight of AI has primarily focused on increasing its effectiveness \cite{sterz2024quest, green2022flaws, enqvist_human_2023, shneiderman_dangers_2016}. In this paper, we highlight the importance of complementing this focus with a security-related perspective on human oversight. This perspective is critical because digital systems are increasingly pressured by cyber criminals and weaknesses in single components can compromise whole IT infrastructures, a risk that will also apply to human oversight as it becomes more widely implemented as part of the safety, security, and accountability architecture of AI operations. Drawing on cybersecurity research \cite{shafahi2018poison,conti2016survey}, we have introduced threat modeling for human oversight. We designed a DFD for the human oversight application, identified assets that need to be protected, and explored vulnerabilities of human oversight using STRIDE. Furthermore, we introduced a minimal set of mitigation strategies to secure human oversight.

Understanding and designing the technical factors, selecting and training human oversight personnel, and shaping working conditions to foster effective human oversight remains an ongoing challenge for research, practice, and AI governance \cite{sterz2024quest, green2022flaws, LauxRuschemeier_2025}. Our analyses indicate that this challenge will intensify as human oversight becomes standard practice in high-risk AI use cases. 
By analyzing an abstracted application-level representation of human oversight using structured threat modeling, we unveiled vulnerabilities that can be exploited by malicious actors to attack key assets, including the requirements of effective human oversight \cite{sterz2024quest}.

Consequently, such attacks may increase risks in AI operations by reducing the effectiveness of oversight, disabling oversight entirely, or manipulating oversight personnel’s intentions to exacerbate risks. We were able to identify a list of security risk using the STRIDE model. However, this list can only serve as a starting point for real applications of human oversight. In practice, our baseline has to be extended with the concrete software and hardware components that are integrated into the human oversight application. These components can open an application to new attack vectors. Furthermore, we included threats posed by misaligned AI agents into our investigation. With growing knowledge about AI agents in practice, new ways possibly emerge how AI agents can deceive oversight \cite{meinke2024frontier,amodei2016concrete}.

While we were using STRIDE to analyze our theoretical human oversight application, additional options are available for the secure implementation of human oversight. Once human oversight is operational, the analysis with STRIDE should be supplemented with operational threat modeling using frameworks like cyber kill chain \cite{yadav2015technical}, MITRE ATT\&CK \cite{al2024mitre}, or unified kill chain \cite{pols2017unified}. This allows for an even more comprehensive and specific analysis of the operational aspects of a human oversight application. Another endeavor that should be considered by practitioners when performing real-world threat modeling is the quantitative analysis of threats. This means determining for each threat the ease of exploitation and the damage potential of an exploitation. Such analysis can help to prioritize hardening strategies.

Identifying risk is only one side of the cybersecurity coin. Appropriate hardening strategies have to be put into place to secure applications. We introduce several hardening strategies that can be used to mitigate the identified vulnerabilities. These strategies are based on the cybersecurity literature  \cite{ross2019developing,reda2023taxonomy,banik2023automated}. However, our provided list has to be understood as a minimal set of measures. Future research needs to identify additional measures and evaluate the effectiveness of common hardening strategies for human oversight. 

Beyond technical hardening, future work could also explore the application of amplified oversight. The core principle of amplified oversight is to employ AI itself to systematically reduce complexity to a level that enables human oversight personnel to fulfill their duties effectively \cite{shah2025approach}. In future, such approaches to support human oversight may even exceed human capabilities and may become a further viable target for malicious actors to attack.

\section{Conclusion}

Research on human oversight has primarily focused on its effectiveness -- and rightfully so, given the complexity of designing oversight processes that genuinely mitigate risks in AI operations \cite{sterz2024quest, laux_institutionalised_2024, green2022flaws}. Adding to this complexity, it will become increasingly important to design for secure human oversight as oversight becomes an integral component of the safety, security, and accountability infrastructure of AI operations. We hope this paper serves as a starting point for research that analyzes the cybersecurity risk of human oversight. Additionally, we hope our paper serves as a guideline for researchers and technicians that are task to implement human oversight in a real-world system.

\section*{Acknowledgments}
The work of Markus Langer is partially funded by the the Daimler and Benz Foundation as part of the project TITAN – \textit{Technologische Intelligenz zur Transformation, Automatisierung und Nutzerorientierung des Justizsystems}  (grant no. 45-06/24). The work of Kevin Baum is partially funded by the German Federal Ministry of Education and Research (BMBF) as part of the project \textit{MAC-MERLin} (Grant Agreement No. 01IW24007), and the European Regional Development Fund (ERDF) as well as the German Federal State of Saarland within the scope of the Project \textit{ToCERTAIN}. In addition, the work of Markus Langer and Kevin Baum is partially funded by the DFG grant 389792660 as part of TRR 248 CPEC --- \textit{Center for Perspicuous Computing}. This work has benefitted from the sixth and last authors' participation in Dagstuhl Seminar Dagstuhl Seminar 25272 "Challenges of Human Oversight: Achieving Human Control of AI-Based Systems".

\bibliographystyle{unsrt}  
\bibliography{references}  

\appendix
\section{Consolidated table of the main results}\label{app:table}
Table \ref{tab:consolidatedTable} consolidates the main results of our paper. It serves as a quick and easy-to-use tool for researchers and practitioners to grasp the most important information and connections presented in the main part of the paper. 

\begin{sidewaystable*}
    \centering
    \caption{Table summarizing the results of the main manuscript. This table is structured around the six groups of threats identifiable using the STRIDE framework. For each threat group, we highlight the parts of the data flow diagram (DFD, Figure 1 in the main manuscript) that are in general susceptible to attacks from the group. The third column provides an overview over the identified assets that are at risk from attacks of each threat group. Finally, we provide a summary of the hardening strategies that can be used to counteract the different threat groups.}
    \hspace*{-1cm}
    \begin{tabular}{lL{6cm}L{6cm}L{6cm}}
        \toprule
        \textbf{STRIDE threat} & \textbf{Affected DFD parts} & \textbf{Assets at risk} & \textbf{Hardening strategies} \\
        \midrule
        Spoofing & Human oversight personnel (correlating to Entry.2), Superior (correlating to Entry.3), External advisory board & Confidentiality (AA.1), Integrity (AA.2), Self-control (AA.6), HO personnel credentials (PA.1) & Training (awareness for social engineering), Encryption, Intrusion Detection System (IDS) \\
        \phantom{placeholder} \\
        Tampering & All data flows, Human oversight IT system & Integrity (AA.2), Epistemic access (AA.4), User data (PA.2) & Transparency (e.g., SBOM for AI), Red Teaming \\
        \phantom{placeholder} \\
        Repudiation & AI system, Human oversight personnel, Human oversight IT system & Confidentiality (AA.1), Integrity (AA.2), Epistemic access (AA.4), Causal power (AA.5), Self-control (AA.6), Fitting attentions (AA.7), Human oversight personnel credentials (PA.1), User data (PA.2) & Red Teaming, Transparency, Training (resilience against coercion, anti-bribery policies) \\
        \phantom{placeholder} \\
        Information disclosure & AI system, Human oversight IT system & Confidentiality (AA.1), Human oversight personnel credentials (PA.1), User data (PA.2) & Encryption \\
        \phantom{placeholder} \\
        Denial of services & AI system, Human oversight IT system & Availability (AA.3), Epistemic access (AA.4), Causal power (AA.5) & Network Management, Intrusion detection system \\
        \phantom{placeholder} \\
        Elevation of privileges & AI system, Human oversight personnel, Human oversight IT system & Causal power (AA.5) & Red Teaming, Transparency \\
        \bottomrule
    \end{tabular}
    \label{tab:consolidatedTable}
\end{sidewaystable*}

\section{Exemplary use case}\label{app:example}
The main manuscript provides researchers and practitioners with a general understanding on how threat modeling in a socio-technical context can be performed. However, moving away from general descriptions to a concrete use case is necessary in translating the generalized and abstract information in actionable insights. This section provides such a use case. First, we will introduce the use case and will show how the different steps of threat modeling are applied to this use case. This will not be an exhaustive threat modeling, since that would result in dozens sometimes hundreds of pages of text and, thus, is not appropriate for a scientific manuscript. Nevertheless, we hope the more hands-on information in this section will be useful for concrete applications of human oversight.

Medical imaging is a common application scenario for AI systems, which are classified as high-risk systems due to the sensitive nature of the personal health data they process and therefore require human oversight under the EU AI Act. The use case presented here describes an AI assistance system in the field of ophthalmology. Optical coherence tomography (OCT), as an imaging technique, is used to diagnose common retinal diseases, such as diabetic retinopathy and age-related macular degeneration. The AI assistance system supports the evaluation of the image data by providing suggestions regarding diagnosis, differentiation between various retinal diseases, and treatment recommendations. The application consists of the following components:
\begin{itemize}
    \item The OCT device, which is a computer with additional hardware parts like a superluminescent diode, different mirrors, and sensors needed for the imaging process.
    \item The IT infrastructure of the medical facility, that includes multiple interface devices for medical staff as well as computing resources to run diagnostic algorithms.
    \item The medical staff needed to perform the OCT measurements and diagnosing patients.
\end{itemize}

\subsection{Step 1 - Scope of the use case}
We will map the DFD shown in Figure 1 in the main manuscript onto our running example. First all the nodes will be put into the context of the medical decision support system. Afterwards we describe what the different data flows mean in this context.

\subsubsection{DFD nodes}
The DFD nodes in the context of our decision support system are defined as follows:
\begin{itemize}
    \item \textbf{Users}: This node represents patients that visit a clinic to be examined. Usually, medical imaging procedures are performed, if there is an indication. Therefore, we can assume that patients have some sort of ailment. Additionally, there is the role duality mentioned in the main manuscript present in our use case. Medical doctors are using the AI system to perform diagnosis and treatment planning, therefore they are considered users as well as oversight personnel in our use case.
    \item \textbf{AI System}: This node represents the whole decision support system. That is not limited to the AI model used for classifying retina scans but also the hardware and software components that are used to generate the images and the software components that are processing the raw imaging data for the medical staff. Arguably, even the hardware that is used to store the imaging data belongs to the AI system and is therefore represented by this node.
    \item \textbf{Human Oversight Personnel}: In case of our decision support system, the human oversight personnel is the medical staff, usually medical doctors, that are responsible for diagnosing patients.
    \item \textbf{Human Oversight IT System}: This node represents all hardware and software components that are used by medical staff to evaluate the AI output. This can include functionalities like additional statistical analyses of the raw image data or the highlighting of parts of the image that played an important part for the AI decision.
    \item \textbf{Superior}: In our use case, this node represents two entities within the clinical environment. First, we have a medical superior that is responsible for professional supervision of medical staff. This is usually the chief physician, chief medical officer, or medical director. Second, there is a economical superior that is responsible for the business side and operational management of a clinic or hospital. This is usually the business director. There is the possibility that these two positions are filled by the same person. This is more likely for smaller clinics.
    \item \textbf{External Advisory Board}: Most countries have established structures that act as advisory and oversight entities for medical facilities operating in these countries. Sometimes these entities are also authorized to publish binding guidelines for diagnosis and treatment of different medical conditions. Our DFD incorporates these structures with the \textit{External Advisory Board} node.
\end{itemize}

\subsubsection{DFD data flows}
For our use case, we can specify the data that is transferred over the data flows drawn in the DFD:
\begin{itemize}
    \item \textbf{Users $\rightarrow$ AI System}: The input that is fed into the AI system by users is relatively restricted in our use case. Patients are providing images of their eyes as input. No other input modality is available to patients. The data send to the AI system by medical doctors are control commands, e.g., requesting certain image slices or processed images.
    \item \textbf{Users $\leftarrow$ AI System}: This data flow is limited for the patient user group in our use case. There could be instructions send from the AI system to the patients. For example, the patient might need to reposition to ensure optimal conditions for the imaging procedure. The AI system could detect such needs and give detailed instructions to the patients.
    Medical doctors will receive images created by the OCT scanner as well as the results of the AI prediction process.
    \item \textbf{Users $\rightarrow$ HO Personnel}: This is a very important data flow in our use case as it is representing the patient consultation process. Patient consultation by medical staff is one of the most important aspects of diagnosis and treatment and also serves as an important oversight signal in our use case.
    \item \textbf{AI System $\rightarrow$ HO Personnel}: Classification results (severeness, location, treatment suggestions); data stream is not exhaustive
    \item \textbf{AI System $\leftarrow$ HO Personnel}: oversight signal to send more/other/additional data from the system to the personnel
    \item \textbf{HO Personnel $\rightarrow$ HO IT System}: Input to display additional information and/or change how the current information is displayed
    \item \textbf{HO Personnel $\leftarrow$ HO IT System}: Processed versions of the raw imaging data and the classification result (raw imaging data: select the correct slice, select the correct ROI, select the correct coloring depending on the severeness; classification result: explanations, additional statistical information) 
    \item \textbf{HO Personnel $\rightarrow$ Superior}: documentation of consultation/diagnosis/treatment process (medical report), ward rounds
    \item \textbf{HO Personnel $\leftarrow$ Superior}: work instructions, supervision
    \item \textbf{External Advisory Board $\rightarrow$ HO Personnel}: binding guidelines regarding consultation/diagnosis/treatment
\end{itemize}

\subsubsection{Entry points}
The first entry point is the patient itself, which will be placed in front of the OCT device and is interacting directly with this component. Another entry point is the interface used by medical staff to perform the OCT procedure. This interface is part of the OCT device. The third entry point into our use case application is the login procedure of the HO personnel. In our use case, this would be the work station used by the medical doctor to examining the OCT results and deciding on treatment options. The fourth possible entry point is via superiors like chief physicians or the medical director with authority to issue instructions. If the application is hosted on the network of the medical facility, there is a fifth entry point via the network.

\subsubsection{Exit points}
If the OCT device has the ability to directly issue instructions to patients, e.g., to improve the posture of patients in order to improve the image quality, this would be the first exit point. However, we do not include this exit point into our use case analysis since such a communication channel is not standard in state-of-the-art OCT devices. Another exit point are data transfer protocols to other medical systems to fulfill documentation requirements or to be used in medical supervision.

\subsubsection{Assets}
The physical assets of our use case application are the login data of medical staff that allow them to conduct the examination, as well as the login data of the physicians used to access the computer used for analysis the OCT output and deciding on treatment options (represented by the HO IT System node in our DFD). Further physical assets are patient data as well as the software and hardware of the proprietary OCT device. 

All of the abstract assets mentioned in the main manuscript apply for the presented use case.

\subsection{Step 2 - Security threats to the use case}
As mentioned in the beginning of the appendix section, we will not exhaustively cover every single aspect of our use case in the STRIDE analysis and will not identify every attack vector. Instead we will provide a concrete example on how STRIDE maps to specific use cases in contrast to the more general view provided within the main manuscript. One concrete threat action is explored for each of the six parts of STRIDE. This approach is meant to provide an glimpse of the vast amount of possible threat actions that have to be covered in a real-world threat modeling. We will repeat the definitions of each part of STRIDE for the convenience of the reader.

\subsubsection{Spoofing} The term spoofing summarizes threat actions that are aimed at accessing and using credentials without permission.

Many modern clinics allow their employees to access the clinic's network remotely via VPN tunnels. In our use case, such a VPN tunnel exists and can be used by the medical staff (i.e., the human oversight personnel) to remotely access the human oversight IT system. This scenario corresponds to access point \textit{Entry.2} (HO personnel login) identified in the main manuscript. Malicious actors can send a manipulated mail to the human oversight personnel using spoofing techniques to make the mail look like it was send from the clinic's IT support. The mail tries to manipulate the receiver in exposing their credentials for the VPN tunnel. Once this attack is successful, attackers can try to log into the clinic's network using the human oversight personnel's credentials. If successful, attacks will most likely have access to the human oversight IT system. Since all retina scans have to be stored in some form in this system, attacks have access to very sensitive information that can be used to cause tremendous damages to patients.

\subsubsection{Tampering} The term tampering summarizes attacks that are aimed at maliciously modifying a system.

In our use case, the patient will be positioned in front of the retinal scanner and several images of the patient's eyes will be recorded. A malicious actor can exploit this setup by wearing contact lenses that are imperceptible for medical staff or the retinal scanner but will change the behavior of the AI model embedded into the retinal scanner. This attack will fall under the field of adversarial attacks and can cause severe security issues with the whole application ranging from simple misclassifications to misaligned behavior that poses a risk to the integrity of the system's functionality and to the confidentiality of sensitive data by the AI model depending on the privileges and the capabilities that the AI model possesses.

\subsubsection{Repudiation} The term repudiation summarizes attacks that are aimed at performing prohibited operations within an IT system.

For our use-case we assume, that part of the HO IT system is a desktop computer that is used by medical stuff to analyze the results of the AI system. Often these desktop computers are located within the same room used for patient consultation. If that is the case, threat actors could use the relatively easy access to these rooms (e.g., it is expected of them to enter the room, of the threat actor is a patient) to modify these computers. A standard tool that can be used is a \textit{rubber ducky}, which is a keystroke injection tool that looks like a USB stick. All such a threat actor has to do is to insert the \textit{rubber ducky} into an open USB socket of the desktop computer and the tampering attack was successful. Since desktop computer are usually placed on-top or below a desk and they often have USB ports at the back, inserting a USB stick without alerting the medical staff is achievable without to much efforts. 

\subsubsection{Information disclosure} The term information disclosure summarizes attacks that aim at reading files without permission or reading data in transit.

As mentioned in the main manuscript, one of the main data flows that could be targeted for these attacks is the data flow between HO personnel and HO IT system. In our use case, that data flow is especially vulnerable when medical staff uses a desktop computer for accessing the HO IT system in the same room as patient consultations take place. Similar to the attack vector sketched out in repudiation, threat actors can get physical access to these computers with relative ease. Once they have physical access to the computer, the whole tool-belt of cybercriminals are available to them to ex-filtrate sensitive information. If the threat actors are left alone in the consultation room, as it often happens in hospitals due to staff shortage, it becomes even easier to perform such attack undetected.

\subsubsection{Denial of services} The term denial of services covers attacks attempting to deny access to valid users, such as by making a web server temporarily unavailable or unusable.

For our running example the threats for the HO application are overlapping with general threats to the hospital's or clinic's IT infrastructure in general. The HO IT system will most likely run on the IT infrastructure located within the medical facility and attacking this infrastructure with a (D)DOS or ransomware attack will also disable the HO IT system. The retinal scanner in our use case is a medical equipment that can also function in isolation, i.e., even if the medical facility's IT infrastructure is disabled. This can allow to perform emergency examinations but effective human oversight of these examinations would not be possible.

\subsubsection{Elevation of privilege} The term elevation of privilege summarizes attacks that aim at escalating the own privileges within a system.

Let's assume that the medical facility utilizes a network-accessible storage (NAS) solution to store all its medical data. The AI system has the right to access a specific location within the NAS with write-only rights to store all images and classification results. Since threat actors disguised as patients can come in physical contact with the AI system, exposed external interfaces (e.g., USB sockets) can be exploited to inject malicious software into the AI system. Combined with exploiting vulnerabilities of the NAS solution, this can result in the elevation of the access privilege of the AI system. Hence, threat actors might be able to access data that are usually inaccessible to the AI system.

\subsection{Step 3 - Countermeasures and hardening strategies for the use case}

All the hardening strategies introduced in the main manuscript apply to our use case. However, we will provide exemplary hands-on hardening strategies that are important to this specific use case. Similar to the attack vectors, this section is far from an exhaustive list. It again serves as a showcase for the variety of options that need to be considered when performing threat modeling on real-world applications. 

In the paper's subsection on security threats, there is one potentially dangerous scenario that needs special attention in terms of hardening strategies. This scenario is the relative ease with which threat actors can get into physical contact with the hardware components of the use case (i.e., the hardware of the AI system and the HO IT system). Therefore, every exposed interface option (e.g., USB sockets, LAN sockets, DisplayPorts) has to be sufficiently secured. A possible option would be a whitelisting of external devices. This process involves creating and maintaining a registry for all devices (like USB sticks) that have the permission to interact with components of the human oversight application. Additionally, all components of the human oversight application need the ability to detect whether a connected external device is on the whitelist. The result is that any device that is connected to components of the human oversight application but does not appear on the whitelist will not be mounted. Thereby, the attack vectors described in the section about repudiation and elevation of privilege are not readily available to threat actors. However, this approach needs a careful balancing of security and usability. On the one hand, the most secure option would be to forbid all external devices to be mounted by the hardware components of our use case. On the other hand, such extreme measures could render the application completely unusable. A secure but still usable middle ground has to be determined.

Another use case specific hardening strategy could be certain organizational measures, e.g., mandatory work guidelines. The medical institution could establish a work instruction that instructs employees to never leave patients unattended with the hardware components of our use case. This would significantly increase the difficulty to secretly insert a malicious devise into an interface option of our application's hardware components. Thereby, hardening the application against some of the attack vectors mentioned in the previous subsection.

As mentioned in the description of the users node, our use case is highly-impacted by the role duality problem. That means, that the human oversight personnel will also be a user of the AI system in our use case. This can easily be verified by noticing that medical doctors are intended to use the predictions of the AI system to support their diagnosis and treatment decisions. Simultaneously, medical doctors are also employed to act as human oversight personnel for the AI system. This role duality can cause conflicts in the oversight structures. Medical doctors (as human oversight personnel) might neglect their oversight duties, if dutifully perform the oversight would result in the disclosure of a medical error by the same medical doctors (as users). This conflict of interests has to be considered in designing hardening strategies for our use case, hence, a multi layered oversight structure has to be implemented. Such a structure can mitigate the risk arising from the role duality problem by decentralizing the oversight structure and distributing power between different parties.

\end{document}